# Enhancing charge-density-wave order in 1T-TiSe$_2$ nanosheet by encapsulation with hexagonal Boron Nitride

L. J. Li[1,2], W. J. Zhao[1,3], B. Liu[1,2], T. H. Ren[1,2], G. Eda[1,2,3], K. P. Loh[1,2]*

[1]Centre for Advanced 2D Materials and Graphene Research Centre, National University of Singapore, 117546, Singapore

[2]Department of Chemistry, National University of Singapore, 117543, Singapore

[3]Department of Physics, National University of Singapore, 117542, Singapore

**Layered transition metal dichalcogenides (TMDs) provide an ideal platform for exploring the effects of dimensionality on correlated electronic phases such as charge density wave (CDW) order. When TMDs are reduced in thickness to the 2-D limit, it is expected that the substrates will exert considerable influence on the electron states. Here we report a study of the charge density wave (CDW) state in 1T-TiSe$_2$ nanosheets of different thicknesses when the sheets are encapsulated by hexagonal Boron Nitride (h-BN) or supported on SiO$_2$ substrate. Our results show that dimensionality reduction results in an enhancement of CDW order and that disorder and substrate phonons tends to destroy CDW order, preventing observation of intrinsic CDW transition in ultrathin samples. Encapsulated 10 nm thick 1T-TiSe$_2$ samples exhibit intrinsic CDW with transition temperature as high as 235 K. Our study points out that choosing the right substrate is important in the search for room temperature CDW materials.**

The discovery of graphene heralded an era of intense research on two dimensional (2D) materials[1]. 2D transiton metal dichalcogenides TX$_2$(T=Mo, W, Nb, Ti; X=S, Se, Te) provide an ideal system for studying many-body interactions, which manifest strongly in 2D systems due to the enhancement of Coulomb interactions and reduced screening[2]. Since 2D materials exhibit very large surface area and low carrier density, the presence of a substrate is expected to influence its electronic properties significantly[3]. In fact, the electron-electron (e-e) and electron-hole (e-h) interactions in graphene can be manipulated by using specific substrates[4, 5]. Generally, to avoid screening of many body interactions, an insulator substrate is the preferred choice for studying the many body effects of 2-D materials. Hexagonal boron nitride (h-BN) has a large band gap, an atomically smooth surface that is relatively free of dangling bonds and charge traps, thus, it is recognized generally to be a disorder-free substrate. Graphene devices fabricated on h-BN show carrier mobility close to that of the free-standing graphene and are superior to those on standard SiO2 substrates[6].

Charge density wave (CDW) is one of the most intensively studied collective quantum states in solid state physics[7, 8]. Due to sharp change of the resistance at the CDW transition temperature $T_{CDW}$, CDW materials show prospects for optoelectronic devices[9, 10] and quantum information processing[11]. For practical applications, CDW materials with a transition temperature near room temperature are desirable. Ultrathin CDW materials offer opportunities for electrically tunable CDW state[12-16] as well as investigating the influence of dimensionality on CDW transition[17-21]. Xu et al., reported increase of $T_{CDW}$ from 100 K for the bulk crystal to 135K for few-layer $VSe_2$ while Xi et al. reported an increase of $T_{CDW}$ from 30 K for the bulk crystal to 150 K for monolayer $NbSe_2$. $T_{CDW}$ in these cases is still far from room temperature. 1T-$TiSe_2$ has a $T_{CDW}$ of 200 K[22] in its bulk form, hence it is a good candidate for attaining room temperature CDW by tuning its dimensionality. Previously, P. Goli et al[17] reported that $T_{CDW}$ increases in 1T-$TiSe_2$ when it is thinned down from few microns to about 50 nm, however thicknesses below 100 nm have only one data point. Furthermore, to date, there is no theoretical or experimental indication (except Ref.17) that dimensional effect exists or is expected since the coherence length of CDW state in TiSe$_2$ is only a few lattice parameters due to strong coupling[23]. The behavior of the CDW state at the quasi-2D limit, where the influence of the substrate gets increasingly stronger, is not known and has not been carefully explored to date.

Herein, we interrogate the CDW state in TiSe$_2$ nanosheets with thicknesses ranging from 3D to the quasi-2D limit (> 100nm to ~ 10nm) using both temperature dependent Raman and resistivity measurements. We found that T$_{CDW}$ of TiSe$_2$ is strongly substrate dependent when the thickness is below 80nm. When supported on SiO$_2$ substrate, T$_{CDW}$ decreases with thickness of TiSe$_2$, showing opposite behavior than that reported in ref. 17, which the possible reason will be discussed in results section. In contrast, T$_{CDW}$ increases as thickness of TiSe$_2$ decreases when the sample is encapsulated in h-BN film. These results indicate that the influence of different substrates on the CDW state is significant when the material goes to quasi 2D. CDW order can be enhanced through both dimensionality reduction and suppression of disorder on substrate. The abnormal dimensionality effect showing in thickness range of few tens of layers needs further theoretical or experimental investigations.

TiSe$_2$ single crystals were grown by conventional chemical vapor transport method[22]. The quality of the crystals was confirmed by X-ray diffraction (supporting information), and its stoichiometry confirmed by energy dispersive X-ray spectroscopy (EDX). Flakes of different thickness were exfoliated and transferred onto silicon substrates with 300nm SiO$_2$ layer in an Ar-filled glovebox. The thickness of the flakes was determined by the contrast of optical microscopy image as well as step height measurement using atomic force microscope (AFM). The encapsulation of h-BN flakes was done by the dry transfer technique[24]. Four-terminal devices for resistivity measurement were fabricated by standard electron beam lithography (EBL) and contacts were made by thermal evaporation of 5 nm chromium and 65 nm gold followed by liftoff. Low temperature resistivity measurement was performed in a physical property measurement system (PPMS-9) from Quantum Design. Lock-in amplifier SR830 from Stanford Research was used for recording the differential resistance. Low temperature Raman measurement was performed in Linkam HFSX350 stage using 633 nm laser with the following parameters (Laser power 0.25mW, step size: 0.33cm$^{-1}$). At each temperature data point, there was a waiting time of 5 minutes for thermal equilibration; thereafter the spectrum was collected using an integrated time of 5 minutes. The estimated temperature difference between cooling and warming process was no more than 5K. Although the flakes being measured have thickness much smaller than 100 nm, any visible damage to the flake or spectral shifts due to local heating problem is not observed.

**Experimental results and discussion:**

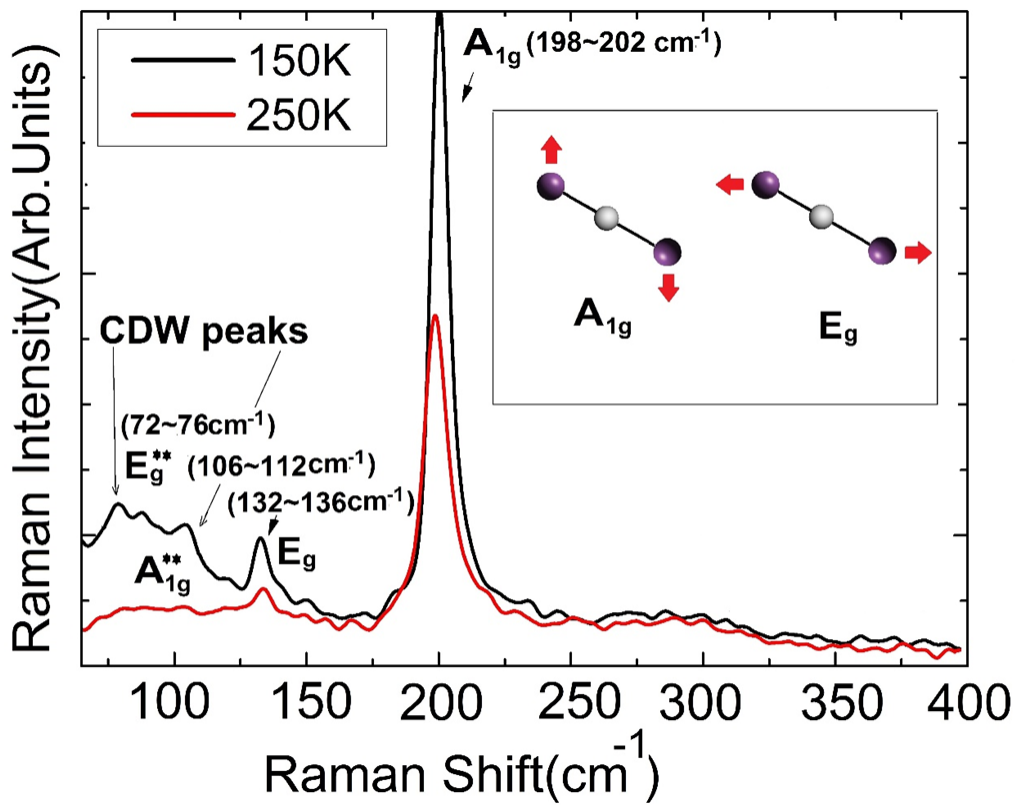

Fig. 1 Raman spectra of 1T-TiSe$_2$, showing the room temperature phase (red) and low temperature CDW phase (black) for exfoliated film with thickness of ~ 21 nm. Inset shows the Raman vibration modes E$_g$ and A$_{1g}$ of 1T-TiSe$_2$. The arrows indicate the direction of vibration.

1T-TiSe$_2$ has a layered structure with the Se-Ti-Se layers separated by van der Waals gap, which allows it to be exfoliated into thin layers. Its lattice space group is hexagonal D$_{3d}^3$. The Ti atom is located in the center of the octahedral, which is the inversion symmetry point. The room temperature phase of 1T-TiSe$_2$ has zero-center vibrational modes[25]:

$$\Gamma = A_{1g} + E_g(2) + 2A_{2u} + 2E_u(2),$$

In which, A$_{1g}$ represents the phonon mode that two Se atoms move along the out-of-plane directions while E$_g$ represents a doubly degenerate phonon mode that the Se atoms move along the in-plane directions, as shown in Fig.1 inset. In Fig.1, the Raman scattering spectra of both room temperature phase and low temperature CDW phase are plotted together. As shown, in both phases, two peaks at 133 cm$^{-1}$ and 201 cm$^{-1}$ are clearly seen, which are close to the E$_g$ and A$_{1g}$ peak positions reported in the previous Raman studies[25, 26]. Nevertheless, in Ref. 17, the authors reported the observation of very weak A1g peak at 199 cm-1 and much stronger Eg peak at 233cm-1, which was completely different from the previous reports and was more likely to be Raman pattern of 2H phase. In accordance with previous observations[25, 26], we observed two peaks related to the CDW phase emerging in the spectrum at 150 K, wherein, the peaks positioned at around 74 cm$^{-1}$ and 110 cm$^{-1}$ are close to the previously reported E$_g^{**}$ mode in bulk TiSe$_2$ at 76 cm$^{-1}$ and A$_{1g}^{**}$ mode at 116 cm$^{-1}$. These two modes are probably associated with the CDW coupled modes from TA phonons at L point[25], where the position shift may come from the thinning effect of the sample and the different equipment offset. Meanwhile, Ref.17 attributed the peak in 133cm$^{-1}$ as the CDW related peak, which was also contradictory to the previous reports.

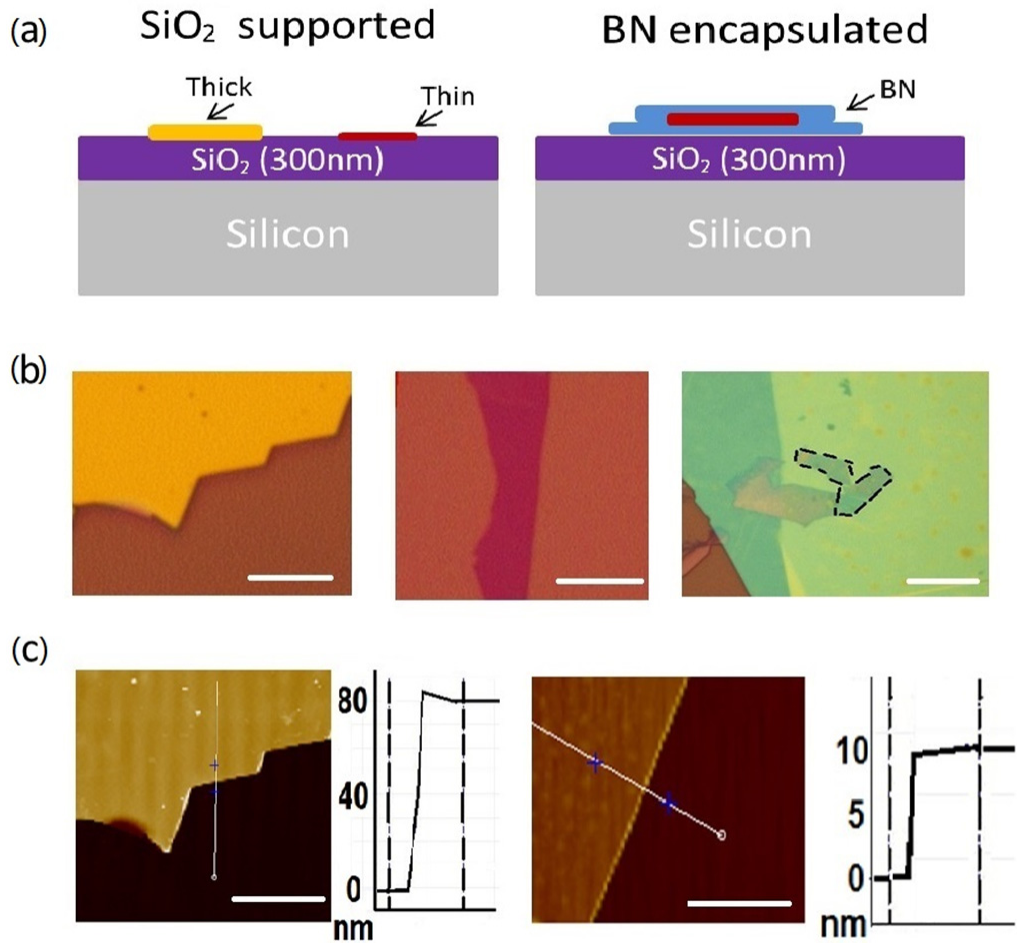

Fig. 2 (a) Schematic pictures of SiO$_2$-supported, and h-BN encapsulated TiSe$_2$ flakes (b) Optical microscope images of the above two types of flakes: SiO$_2$ supported 80 nm(left), 10 nm(middle) and h-BN encapsulated 10nm(right). (c) The AFM image and thickness determination for 80 nm and 10 nm flakes. All scale bars stand for 10 um.

To investigate the substrate effect, we carried out micro-Raman measurement comparatively on SiO$_2$-supported, h-BN encapsulated TiSe$_2$ flakes as a function of thicknesses and temperatures. Fig. 2 displays the schematic, optical microscopy and AFM pictures of the measured samples. Significant differences in the fine peak features were found between the encapsulated flakes and the SiO$_2$–supported flakes and between samples with different thicknesses.

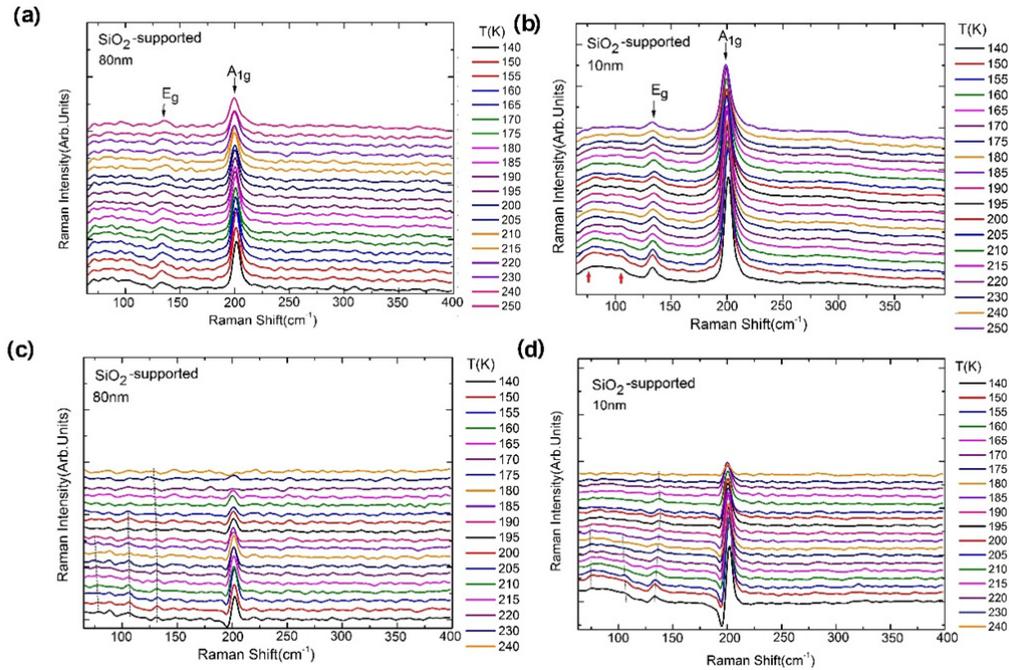

Fig. 3 Temperature dependent Raman spectra of SiO$_2$-supported TiSe$_2$ flakes with thickness of (a) 80 nm (b) 10 nm; the two red arrows indicate the two peaks related to CDW modes (c), (d) show the spectra subtracted by the spectral of 250 K in (a) and (b), respectively. The dotted lines and frames indicate that the peaks emerge and the E$_g$ mode changes between hardening and softening at T$_{CDW}$.

Fig. 3 displays the temperature dependent Raman spectra of SiO$_2$-supported TiSe$_2$ flake with thickness of (a) 80nm and (b) 10 nm, respectively. As shown, the two main Raman active modes E$_g$ and A$_{1g}$ located at 134 cm$^{-1}$ and 198 cm$^{-1}$ respectively can be seen clearly at all the temperatures. At low temperatures, the CDW associated peaks around 74 cm$^{-1}$ and 110 cm$^{-1}$ emerge as shoulder-like humps, as indicated by the two small red arrows. To improve the signal-to-noise ratio, we normalize the spectra by subtracting high temperature spectra (temperature above T$_{CDW}$) from the other low temperatures spectra. This will allow us to remove the temperature independent effect which is not CDW related. Herein, we subtract the spectra of 250 K from all other lower temperature spectra and plot the subtracted spectra in Fig. 3(c) and Fig. 3(d) respectively, for $SiO_2$-supported 80 nm and 10 nm flakes. After normalization, one can see peaks at around 74 $cm^{-1}$ and 110 $cm^{-1}$ develop at lower temperatures, which are related to CDW order. For $SiO_2$-supported 80 nm, we can also observe the other two $E_g$ peaks reported previously in Ref. (24,25) at around 93 $cm^{-1}$ that are hardly seen in the original spectra Fig. 3(a). However, these peaks are not seen for $SiO_2$-supported 10 nm even in processed spectra Fig. 3(d), which is probably due to the destructive phonon scattering effect from the $SiO_2$ substrate. Specifically, as guided by the dotted arrows, when temperature decreases, the 135 $cm^{-1}$ peak first shifts to higher wavenumbers and then reverses its shift. Judging from the emergence of CDW related Raman peaks by the increasing peak intensity and the shift between hardening and softening of the $E_g$ peak, one can determine the $T_{CDW}$ for the $SiO_2$-supported 80 nm and 10 nm flakes to be around 200 K and 190 K with an error of ± 5 K, respectively (Details of determination of $T_{CDW}$ is shown in supporting information section 2.).

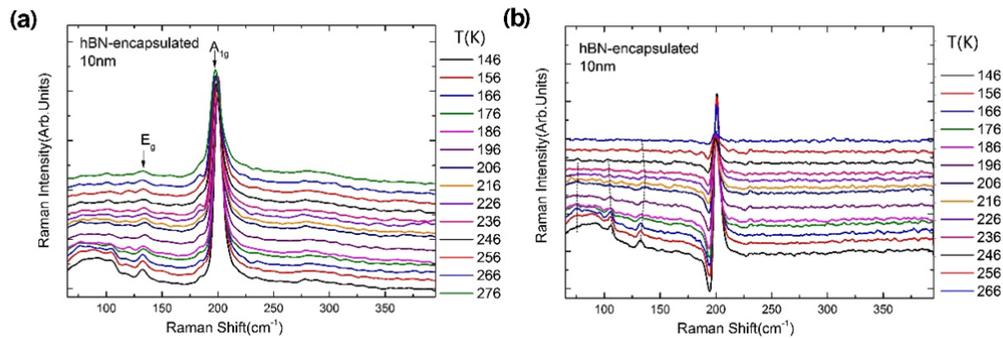

Fig.4 The temperature dependent Raman spectra of h-BN encapsulated $TiSe_2$ flakes with thickness of ~ 10 nm. (a), (b) show the original and the difference spectra subtracted by spectra of 273 K respectively.

To study the influence of the substrate on the CDW state in thin $TiSe_2$ flakes, we performed the same experiments on h-BN encapsulated flakes. Fig. 4 shows the temperature-dependent Raman spectra of an h-BN encapsulated 10 nm-TiSe$_2$ flake. The T$_{CDW}$ are determined again by generating the difference spectra using the same procedure described above. Similarly, one can see clearly the emergence of the two peaks at around 74 cm$^{-1}$ and 110 cm$^{-1}$ and the hardening/softening of the E$_g$ peak at T$_{CDW}$ in the subtracted spectra. Remarkably, T$_{CDW}$ of encapsulated samples were found to be 235 K, which are ~ 40 K higher than that of a SiO$_2$-supported flake. We also performed temperature dependent Raman spectroscopy for other thickness flakes of both SiO$_2$-supported and encapsulated ones (see supporting information). In Fig. 5(a), we summarize the T$_{CDW}$ values derived from low-T Raman spectroscopy for SiO$_2$-supported, and h-BN encapsulated TiSe$_2$ flakes of different thicknesses. The results clearly indicate the enhancement of CDW state for encapsulated TiSe$_2$ nanosheets, compared to SiO$_2$-supported sample of the same thickness.

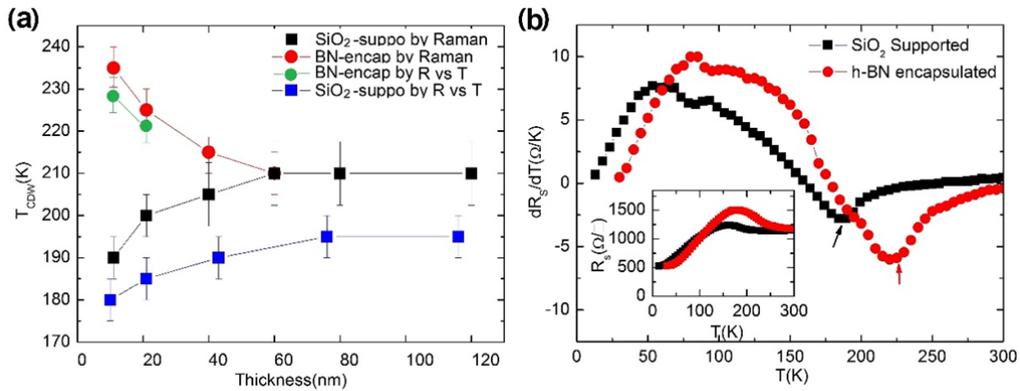

Fig. 5(a) Summary of T$_{CDW}$ derived from the temperature dependent Raman and resistance data for SiO$_2$-supported and h-BN encapsulated flakes. (b) T$_{CDW}$ determined from the inflection point of R (resistance) *vs* T (temperature) (shown in inset) by plots of dR/dT for encapsulated and SiO$_2$-supported TiSe$_2$ flakes of about 10 nm thickness.

Resistivity measurements were also performed to study the dependence of CDW transition on different substrates. The inflection point on the resistivity versus temperature curve is considered to be the onset of CDW formation because of the second order nature of the CDW phase transition. We measured the temperature dependent resistivity for both the SiO$_2$–supported and h-BN encapsulated TiSe$_2$ flakes of different thicknesses. Fig. 5(b) displays the sheet resistance versus temperature curves for flakes of 10 nm thickness. As can be seen from the plot of the dR/dT curves, the T$_{CDW}$ of h-BN encapsulated flakes is about 30 K higher than that supported on SiO$_2$. The significant difference of T$_{CDW}$ between the SiO$_2$-supported flake and the h-BN encapsulated flake is consistent with the result derived from low temperature Raman spectroscopy, which confirms the influence of the substrate on the properties of 2D material. The T$_{CDW}$ derived from R vs T data is summarized in Fig.5(a), more original data and related discussion can be found in the supporting formation. The resistance determined T$_{CDW}$ have a similar temperature dependence to the Raman determined T$_{CDW}$ but are slightly lower, perhaps because the resistance is dominated by the layers closest to the substrate.

With decreasing thickness of the samples, the two Raman peaks (E$_g$ 74 cm$^{-1}$ and A$_g$ 110 cm$^{-1}$) corresponding to CDW order have different strength. The out-of-plane A$_g$ mode peak is noticeably stronger than the in-plane E$_g$ mode peak. This can be explained by the fact that the CDW vector of 1T-TiSe$_2$ is (0.5, 0.5, 0.5) **a**[25]. It has a projection along *c* axis (the out-of-plane direction) so that this vector is easily affected by the out-of-plane phonon interaction, which is accentuated by the reduction in dimensionality and the substrate phonons, while the in-plane phonon mode remains reasonably unaffected by these two effects. This observation is also consistent with previous study on MoS$_2$ [27], which suggest that the out-of-plane A$_{1g}$ mode is more affected by SiO$_2$ substrate than the in-plane E$_g$ mode. This indicates that the phonon or electron-phonon interaction is important in the CDW formation.

There are two main mechanisms under debate for the formation of CDW in TiSe$_2$, namely electron-phonon coupling band-type Jahn-Teller transition[28, 29] and pure carrier dominated excitonic insulator transition[30, 31]. In terms of the latter mechanism, a carrier concentration level of $10^{14}$ cm$^{-2}$ is needed to induce the transition of the CDW state[32]. However, the doping carrier concentration induced by trapped molecules on SiO$_2$ surface like H$_2$O, O$_2$ etc. are usually in the range of $10^{12}$ cm$^{-2}$ (Ref. 6,33). Furthermore, for flakes with thicknesses up to few tens of nanometres, the carrier screening effect is large and cannot be excluded. Therefore, once again we can conjecture that phonon or electron-phonon coupling must play a major role in such large T$_{CDW}$ enhancement instead of the pure carrier doping effect.

**Conclusion:**

In conclusion, we have studied the effects of substrate and reduced dimensionality on CDW transition temperature of TiSe$_2$ using low-T Raman microscopy and electrical measurements. It is found that the CDW transition temperature of ultrathin TiSe$_2$ can be significantly influenced by substrate. H-BN encapsulated TiSe$_2$ flakes show increased T$_{CDW}$ (from 190K to 235K) with a decrease in thickness while SiO$_2$-supported flakes show opposite thickness dependence. Our results reveal that both reducing dimensionality of the CDW material and suppressing disorder on substrate are important strategies to strongly enhance the CDW state, thus pointing the direction forward in the search for room temperature CDW materials.

**Supplementary material**
See supplementary material for the XRD data of TiSe2 single crystal, Raman spectra analysis details, and temperature dependent Resistance data for SiO2 supported samples.